\documentclass[12pt]{article}
\usepackage{amsmath,amssymb,graphicx,cite}
\usepackage{tabularx}
\usepackage[linktocpage]{hyperref}
\hypersetup{
	colorlinks=true,
	linkcolor=blue,
	filecolor=magenta,
	urlcolor=cyan,
	pdftitle={Sharelatex Example},
	bookmarks=true,
	citecolor=blue,
}
\usepackage{geometry}
\usepackage{setspace}
\usepackage{fancyhdr}
\newcommand{\be}{\begin{equation}}
	\newcommand{\ee}{\end{equation}}
\newcommand{\bea}{\begin{eqnarray}}
	\newcommand{\eea}{\end{eqnarray}}

\def\({\left(} \def\){\right)}

\newgeometry{margin=.85in,top=1.5in}
\begin{document}
	
	\renewcommand{\baselinestretch}{1.5}\normalsize

	\title{\vspace{-1.8in}
		{ Discovering Love numbers through resonance excitation during extreme mass ratio inspirals }}
	
	\author{\large Shani Avitan, Ram Brustein and Yotam Sherf
		\\
		\vspace{-.5in}  \vbox{
			\begin{center}
				$^{\textrm{\normalsize
						Department of Physics, Ben-Gurion University,
						Beer-Sheva 84105, Israel}}$
				\\ \small 
				shavi@post.bgu.ac.il,\ ramyb@bgu.ac.il,\ sherfyo@post.bgu.ac.il
			\end{center}
	}}
	\renewcommand{\baselinestretch}{1.35}\normalsize
	\date{}
	\maketitle
	\begin{abstract}	
\noindent
General Relativity predicts that black holes do not possess an internal structure and consequently cannot be excited. This leads to a specific prediction about the waveform of gravitational waves which they emit during a binary black hole inspiral and to the vanishing of their Love numbers.  However, if astrophysical black holes do possess an internal structure, their Love numbers would no longer vanish, and they could be excited during an inspiral by the transfer of orbital energy. This would affect the orbital period and lead to an observable imprint on the emitted gravitational waves waveform. The effect is enhanced if one of the binary companions is resonantly excited. We discuss the conditions for resonant excitation of a hypothetical internal structure of black holes and calculate the phase change of the gravitational waves waveform that is induced due to such resonant excitation during intermediate- and extreme-mass-ratio inspirals. We then relate the phase change to the electric quadrupolar Love number of the larger companion, which is resonantly excited by its smaller companion. We discuss the statistical error on measuring the Love number by LISA and show that, because of this phase change, the statistical error is small even for small values of the Love number. Our results provide a strong indication that the Love number could be detected by LISA with remarkable accuracy, much higher than what can be achieved via tidal deformation effects. Our results further indicate that resonant excitation of the central black hole during an extreme- or intermediate-mass-ratio inspirals is the most promising effect for putting bounds on, or detecting, non-vanishing tidal Love numbers of black holes.

	\end{abstract}
	\newpage
	\renewcommand{\baselinestretch}{1.5}\normalsize

\section{Introduction}	

General Relativity predicts that black holes (BHs) do not possess an internal structure. They are ``bald'' and can be characterize solely by their mass and angular momentum \cite{Bekenstein:1971hc}. Coalescing BHs radiate gravitational waves (GWs) which are being detected by  the  LIGO and VIRGO observatories since September 2015 \cite{LIGOScientific:2016aoc}. The calculational efforts for improving the accuracy of the general relativistic (GR) predictions for the emitted GW waveform, could hopefully provide an opportunity for testing the baldness of BHs. Particularly, the inclusion of tidal interactions may allow us to probe the hypothetical interior structure of the binary companions and quantitatively test the predictions of GR \cite{Flanagan:2007ix,Hinderer:2009ca,De:2018uhw,Maselli:2013mva,Yazadjiev:2018xxk,Pacilio:2021jmq}.
	
	
In spite of  the increasing precision of ground-based detectors, their limited frequency band enables observations of only a few cycles in the inspiral phase of a binary-BH (BBH) coallesence event  for a limited range of masses. The LISA space detector \cite{LISA:2017pwj,LISA:2022kgy}, whose design sensitivity is maximal in the mHZ region, is expected to be able to detect and track  many  BBH coalescence events from the early stages of the inspiral through the merger to the late post-merger ringdown.


In GR, the interior of  BHs is vacuum, except for a possibly singular core. But is this their true description?   A common expectation is that quantum effects at the Planck length scale, or at a somewhat larger length scale, such as the string length scale, will be sufficient to resolve all singularities. However, there are strong indications to the contrary when it comes to the resolution of BH singularities. First, a seemingly necessary condition  for evading the singularity theorems \cite{PenHawk1,PenHawk2} and the closely related  ``Buchdahl-like'' bounds \cite{Buchdahl,chand1,chand2,bondi} is that  the geometry is sourced by matter that has the maximal negative radial pressure permitted by causality, $p_r=-\rho$, all the way to the surface of the star \cite{bookdill}. Furthermore, if one also considers the emitted Hawking radiation from such a quantum-regularized BH, one finds  an untenable violation of energy conservation: When the scale of resolution is parametrically smaller than that  of the Schwarzschild radius, the emitted energy of Hawking particles will greatly exceed the original mass of the collapsing matter \cite{frolov,visser}. Thus, the tentative conclusion that we will adopt in our following discussion, is that deviations from GR must extend throughout the object's interior, that is, horizon-scale deviations from GR.
	
	The Love numbers of an object encode its response to an external tidal field. These numbers could provide some limited information about the mass distribution and the compactness of the object. The Love numbers determine the possible corrections to the GW signal due to tidal interactions of the companion in a binary system. The quadrupolar Love number $k_2$ identically vanishes for GR BHs in four spacetime dimensions \cite{ Damour:2009vw,Binnington:2009bb,Chia:2020yla,Charalambous:2021kcz,Charalambous:2021mea,Hui:2021vcv,BenAchour:2022uqo}, making it a key observable. Measuring non-zero values will indicate a deviation from the GR predictions \cite{Maselli:2017cmm,Cardoso:2017cfl,qLove,cLove,Brustein:2021pof,Datta:2021hvm}. If indeed horizon scale deviations from the GR predictions occur, then the expectation is that the Love numbers will be small, but not extremely small, suppressed only by some additional pertrubative parameter that quantifies the strength of the deviations. The reason for such expectation is that the Love numbers are normalized such that they are order unity if all the dimensional scales are of order of their radius \cite{qLove,cLove}.

Previous studies have primarily focused on measuring the Love numbers using tidal deformability, which constitutes a subleading correction to the emitted GW waveform and enters at 5PN order compared to the dominant point-particle term. Tidal-deformability effects are more pronounced at the late inspiral phase. This makes the measurement of the Love number more challenging, since other finite-size effects are also of similar magnitude, requiring the construction of more accurate GW waveforms and detectors with better sensitivity.
 \cite{Flanagan:2007ix,Damour:2012yf,LIGOScientific:2017vwq,Sennett:2017etc,Piovano:2022ojl}.

For GR BHs the inspiral evolution is dominated by the point-particle GW emission. For BHs which posses an internal structure, an interesting different effect can dominate the evolution if the orbital frequency becomes comparable to a characteristic frequency of some internal mode.
In this case, this mode is resonantly excited, resulting in a rapid energy transfer from the orbit to the internal mode. The loss of orbital energy effectively advances the inspiral evolution, bringing the companions to a closer and faster orbit. The abrupt energy transfer changes significantly the emitted GW waveform compared to the point particle waveform since it leads to an instantaneous phase jump and a secondary subleading accumulated dephasing due to the differences in orbital velocities.
Such resonant energy transfer can only be realized when the internal modes are non-relativistic. The reason is that the Keplerian orbital frequency is much smaller than the relativistic frequency $c/R$ ($c$ is the speed of light and $R$ is the radius of the compact object) when the two objects are far from each other.

Tidal resonant excitations were first discussed in the context of ordinary polytropic stars \cite{Cowling:1941}, then, much later, for binaries with at least one of the companions being a neutron star \cite{ Lai:1993di, Reisenegger and Goldreich,Shibata,Kokkotas:1995xe,Ho,Flanagan:2006sb,Lai:2006pr, Hinderer:2016eia,Berry:2016bit, Yu:2016ltf, Pratten:2019sed,Bonga:2019ycj,Gupta:2021cno}. In these cases,  the effect was related to the tidal Love numbers \cite{Gupta:2023oyy, Gupta:2022qgg, Steinhoff:2016rfi,Chakrabarti:2013xza, Chakrabarti:2013lua,Andersson:2019ahb}.  However, as already mentioned, since the corrections enter the GWs waveform at 5PN order, the effect becomes significant during the late inspiral phase, where additional effects are also significant, making it difficult to detect the Love number with high confidence.

 More recent studies related to BH mimickers \cite{Cardoso:2019nis,Asali:2020wup,Fransen:2020prl, Maggio:2021uge}, treat the BBH as if they were horizon-less ultra-compact objects (UCOs). In \cite{Cardoso:2019nis,Fransen:2020prl,Maggio:2021uge}, the tidal field was exciting some additional spacetime modes of a hypothetical spacetime structure outside the UCO. In \cite{Fransen:2020prl}, the resulting phase shift due to the resonant excitation  of these additional spacetime modes was related to the tidal Love numbers, and the detectability of the quadrupolar Love number $k_2$ using observations of ground-based GW detectors and the proposed Einstein telescope was discussed. In \cite{Asali:2020wup}, the detectability prospects of the resonance effects were discussed, but  without connecting the effect to the tidal Love numbers. In this study, no evidence for resonance was found in the observations of the first two runs of Advanced LIGO and Advanced Virgo.

 Here, in contrast to previous studies, we discuss the tidal excitation of hypothetical non-relativistic internal modes of the BH, relate the resulting phase shift to the Love numbers and discuss the possible detectability of $k_2$ in LISA observations of IMRIs and  EMRIs. We find that this is, by far, the most promising way to put bounds on, or detect Love numbers of astrophysical BHs.

	
In the following, contrary to GR, we assume that astrophysical BHs do have an internal structure that supports non-relativistic fluid modes.
We keep the calculations as model–independent as possible by expressing the model-dependent quantities through the dimensionless tidal Love number.
We follow the discussion in  \cite{Fransen:2020prl,qLove}, to relate the resonance phase shift of the excited modes to the quadrupolar tidal Love number $k_2$, and their relation to internal modes frequencies of quantum black holes \cite{collision,qLove,cLove,Brustein:2021pof} and recent frozen star model results \cite{FS1,FS2,FSpert}.
We estimate the statistical error in the measurement of $k_2$ through resonance excitations during the inspiral of slowly rotating EMRIs and IMRIs, using the design noise spectrum of LISA \cite{LISA:2017pwj, Robson:2018ifk}. We find that the  statistical error is small even for small values of the Love number, providing a strong indication that the Love number could be detected with impressive accuracy.
We end with an explicit comparison between the detection prospects of the Love numbers with tidal deformability and tidal resonance, and conclude that resonance excitations are the most promising effect for  detecting the Love numbers.
	
\section{Tidal-Resonance interaction }
		
Here, we examine the tidal interaction in a binary system, focusing on the central object that is subjected to the weak periodic tidal force exerted by the smaller companion. Following the ideas presented in \cite{Lai:1993di,Ho,Chakrabarti:2013xza,Andersson:2019ahb}  and more recently in \cite{cLove}, we describe the response of the object to the tidal force from the perspective of an asymptotic observer. The idea is that the object possess a set of  non-relativistic fluid modes which are driven by the tidal force and can be therefore described as a collection of driven harmonic oscillators.

The spectrum of the interior fluid modes depends on the radial number $n$ and the angular numbers $l,m$, so their frequencies depend of the three numbers $\omega_{nlm}$. Here, we are particularly interested in the dominant effect which is due to the excitation of the $n=1$ mode by the quadrupolar tidal field, so we focus on the case $l=m=2$ \cite{Ho}. As for the other modes; the spherically symmetric static $m=0$ mode cannot generate pressure gradients that are needed for resonance excitaion and therefore is not relevant to our discussion. The $m=1$ mode can be resonantly excited in the case that the spin-orbit configuration is misaligned \cite{Ho,Lai:2006pr}. Here, we restrict our attention to spin vectors that are aligned with the orbital angular momentum.

The mode corresponding to $n,~l,~m = 1,~ 2,~ 2$ is non-relativistic, meaning that, as for neutron stars, $\omega_{122}$ is parametrically smaller than $c/R$. The orbital frequency which determines the frequency of the driving tidal force is determined by the Kepler law. It follows, as explained in the Introduction, that as the smaller object gets closer to the central object, the orbital and internal frequencies can match.

When the  frequency of one of the interior modes of the central, larger, object, matches the orbital frequency of the companion, it is resonantly excited and efficiently absorbs energy from the orbital motion. The instantaneous energy absorption increases the orbital velocity and shortens the inspiral duration, thus leading to a phase difference in the emitted GW waveform, when compared to the emitted waveform  in the absence of a resonance. To calculate the dephasing of the GW waveform, we adopt the derivation in \cite{Reisenegger and Goldreich,Flanagan:2006sb},  resulting in the following phase evolution,
		\begin{gather}
			\begin{cases}
				\Phi(t)=\Phi_{PP}(t) & t < t_R\\
				\Phi(t)=\Phi_{PP}(t+\Delta t)-\Delta\Phi & t>t_R+\Delta t,
			\end{cases}
			\label{rp1}
		\end{gather}
where $\Phi_{PP}(t)$ is the point particle phase, $t_R$ is the time at which the resonance starts, $\Delta t$ is the resonance duration and $\Delta \Phi$ is the instantaneous resonance phase difference, which in general depends on the object's properties as demonstrated below. The point particle phase $\Phi_{PP}$, is independent, by definition, on the object's composition. In particular, it has the same value for a GR black hole and one endowed with an internal excitation spectrum, such as the objects we are discussing.
		
Then, assuming that the resonance duration is short compared to the inspiral duration  and under adiabatic evolution, we arrive at the frequency domain resonance phase \cite{Reisenegger and Goldreich,Flanagan:2006sb},
		\begin{gather}
			\Phi(f)~=~\Phi(f)_{PP}+\Theta(f-f_R)\left(\dfrac{f}{f_R}-1\right)\Delta \Phi_{Res}~,
			\label{pf}
		\end{gather}
where $f_R$ is the internal mode frequency which satisfies the resonance condition $2\pi f_R=m\Omega$, $\Omega$ being the orbital angular velocity. Resonance corrections to the phase $\Delta \Phi_{Res}$, are composed of two terms; a dominant term that enters at 2.5PN order higher than the leading order point-particle term and a subleading 4PN-higher contribution. The dominant contribution, which is frequency independent and proportional to $\Delta  \Phi$, originates from the instantaneous energy absorption during resonance. The subleading term, which is proportional to the frequency, is a secular effect that increases towards the late stages of the inspiral.
		
\subsection{The phase shift}	

Fluid perturbations of compact objects are described by the displacement vector $\xi^i$, of a fluid element from its unperturbed position, which is given by the orthonormal base decomposition,
		\begin{gather}
			\xi^i ~=~\sum_n a_n \xi^i_n,
		\end{gather}
$\xi_n$ being the normal displacement vectors, and $a_n$ are the dimensionless displacement amplitudes.\footnote{We use relativistic units $G,c=1$.}
		In the presence of tidal forces, the fluid modes satisfy the damped-driven harmonic oscillator equation \cite{Lai:1993di,Lai:2006pr},
		\begin{gather}
			\ddot{a}_{nlm} +2\gamma_{n}\dot{a}_{nlm}+\omega_n^2 a_{nlm}~=~\mathcal{F}(t)_{nlm},
			\label{eom1}
		\end{gather}
where  $\gamma_n=-\text{Im} ~\omega_n$ is the damping rate of the mode. The source of the damping and its precise magnitude are irrelevant for the resulting resonant excitation and dephasing. So, $\gamma$ can be neglected altogether (see below).

The external periodic force $\mathcal{F}(t)_{nlm}$ excites the $n$th mode interior fluid mode is  given by
		\begin{gather}
			\mathcal{F}(t)_{nlm}~=~N_{lm}{\dfrac{\mathcal{E}_lQ_{nl}	}{MR^2}e^{-i m \phi (t)}}~,
		\end{gather}
where $M$ and $R$ are the mass and radius of the central object.
The order unity factor  $N_{lm}$  is proportional to the Wigner function and is specified below. The tidal field of the $l$ mode is denoted by $\mathcal{E}_l$,  which for the $l=2, m=\pm2$ satisfies $\mathcal{E}_{ij}x^ix^j=\mathcal{E}r^2Y_{2\pm2}$.
The mass moment of the quadrupolar $n$th mode $Q_{n}$  is given by the overlap integral \cite{Reisenegger and Goldreich},
		\begin{gather}
			Q_{n}=-\int d^3r\delta\rho_n r^{2} ~,
\label{qnl}
		\end{gather}
where $\delta\rho_n$ is the corresponding energy density perturbation.

Next, we aim to find the instantaneous phase shift $\Delta \Phi$ and the corresponding phase evolution in Eq.~(\ref{rp1}). We start by solving Eq.~(\ref{eom1}) for the amplitudes $a_n$, which at resonance is given by \cite{Lai:2006pr},
		\begin{gather}
			a_n(t)~=~	\left({\dfrac{\pi}{m\ddot{\phi}}}\right)^{1/2} \frac{\mathcal{F}(t)_{nlm}}{\gamma_{nl}-i\omega_{nl}}e^{-i\omega_{nl}t},
			\label{amp1}
		\end{gather}
where $\ddot{\phi}$ denotes the rate of change of the orbital frequency at resonance.
The transferred energy to the mode $nlm$ during the resonance is a sum of kinetic and potential terms \cite{Lai:1993di,Lai:2006pr},
		\begin{gather}
	E_{nlm}(t)~=~\left(\dfrac{1}{2}\dot{a}_{nlm}(t)^2+\dfrac{1}{2}\omega_{nl}^2a_{nlm}^2(t)\right)MR^2~.
		\end{gather}
	The total energy absorbed by the mode, neglecting $\gamma_{nl}$, is given by
	\begin{gather}
\Delta E_{nlm}~=~~N_{lm}^2\dfrac{\pi }{4 m \ddot{\phi}}\dfrac{(\mathcal{E}_{l}Q_{nl})^2}{MR^2}
\label{ea1}~.
	\end{gather}

The resonance excitations lead to a phase shift, since the orbital energy decreases as it excites the interior modes. Accordingly, the orbital velocity increases and the inspiral duration decreases by a time $\Delta t$. To estimate $\Delta t$, we follow \cite{Flanagan:2006sb}.  The energy absorbed by the central objects decreases the energy of the orbit by the same amount.  In the absence of resonance, such a decrease in energy can only occur by the emission of GW and the time that it would take the orbit to emit GW with such energy  $\Delta t$  would be determined by the equality $\dot{E}_{GW} \Delta t =\Delta E_{nlm}$. The rate of GW emission $\dot{E}_{GW}$ is, to a very good approximation, the same rate as in the absence of resonance, which to leading order is given by $\dot{E}_{GW}=\frac{32}{5}(\mathcal{M}_c~ \Omega)^{10/3}$, with $\mathcal{M}_c$ being the chirp mass. The resulting phase shift $\Delta \Phi=m\Omega \Delta t $ is the following,
		\begin{gather}
			\Delta \Phi_{nlm}~=~m\Omega\dfrac{ \Delta E_{nlm}}{\dot{E}_{GW}}=\frac{5}{32}~m\Omega~\dfrac{ \Delta E_{nlm}}{(\mathcal{M}_c ~\Omega)^{10/3}} .
		\end{gather}	
For IMRIs or EMRIs $\mathcal{M}_c\approx M$ and $\dot{E}_{GW}\sim v^{10}$.

		Using Eq.~(\ref{ea1}), we may calculate the phase shift induced by  the leading order quadrupolar mode $l=m=2$ \cite{Reisenegger and Goldreich,Fransen:2020prl},
		\begin{gather}
			\Delta \Phi_{n22}~=~\dfrac{25\pi }{1024 q(1+q)}\dfrac{1}{ R_1^5}\dfrac{|Q_{n2}|^2}{M_1\omega_{22}^2R_1^2}= \dfrac{25\pi }{2048 q(1+q)}\dfrac{1}{ R_1^5}\dfrac{|Q_{n2}|^2}{\Delta E^{int}},
\label{phasemain}
		\end{gather}
where we used that $N_{22}=\sqrt{3/32}$. Here $q=M_2/M_1$ is the mass ratio and $\Delta E^{int}=\frac{1}{2}M_1\omega_{22}^2R_1^2$ is the internal energy of oscillations which is related to the energy stored in the $n$th mode by $\Delta E^{int}=\sum\limits_n \Delta E_{n22}$, \cite{Chakrabarti:2013xza}.

We wish to justify our estimate of $\Delta t$ using only $\dot{E}_{GW}$ and neglecting other dissipation effects. In general, the time difference $\Delta t$ should include all types of dissipation channels, mainly the dominant dissipation due to tidal friction and the subleading tidal deformation. However, the rate of work of tidal friction is given by \cite{Poisson:2009di,Sherf:2021ppp} $\dot{E}_{TF}=\frac{1}{2}Q_{ij}\dot{\mathcal{E}}^{ij}\sim k_2v^{15}\nu/M$, where $\nu$ is the kinematic viscosity giving rise to viscous dissipation. In \cite{Sherf:2021ppp}, it is demonstrated that, under reasonable assumptions,  the contribution of viscous dissipation is negligibly small compared to the leading order GW emission and, therefore, can be ignored. For example, for cold Neutron stars, considered to be highly viscous $\nu/M\approx 10^{-7}$, whereas for BHs $\nu/M=1$ \cite{Thorne:1986iy}. During the inspiral, when the orbital velocity is non-relativistic the ratio of the different emission rates scales as $\dot{E}_{TF}/\dot{E}_{GW}\sim v^5\ll 1$, which shows that the internal dissipation effects can indeed be neglected.

\section{Fluid-origin Love numbers}

Here we follow \cite{qLove,cLove} to determine the relationship between the Love number and the spectrum of internal fluid modes. We focus on the static tidal Love number, ignoring dissipative effects.

Following \cite{cLove} (see also \cite{Andersson:2019ahb, Chakrabarti:2013xza}), we wish to find the static response of the object to an external tidal field. At low frequencies, away from resonance, the amplitude in Eq.~(\ref{amp1}) reduces to
 \begin{gather}
a_n=\dfrac{\mathcal{E}Q_n}{M\omega_n^2R^2}~.
\label{amplove}
 \end{gather}

 Then, using the definition of the Love number, $k_2R^5=3Q/(2\mathcal{E})$, we apply the normal mode decomposition identities $Q=\sum_n a_n Q_n$, and  $k_2=\sum_n a_n k_{2n}$, where the n$th$ mode Love number, which is associated with the n$th$ mode quadrupolar moment, is given by
   \begin{gather}
k_{2n}R^5=\frac{3 Q_n}{2\mathcal{E}}~.
\label{loven}
   \end{gather}
when substituting the explicit form of $a_n$ from Eq.~(\ref{amplove}), the Love number becomes
 \begin{gather}
 	k_2~=~\sum_n \dfrac{3}{2 R^5}\dfrac{Q_n^2}{M \omega_n^2 R^2}~.
 \label{k21}
 \end{gather}

We now approximate $k_2$ by the first term in the sum in Eq.~(\ref{k21}) relying on a physically motivated assumption. The sum in Eq.~(\ref{k21}) is dominated by the fundamental $n=1$ mode. The justification is that the number of nodes in the overlap integral in Eq.~(\ref{qnl}) increases as $n$ increases. It follows that the contribution of $Q_n$ decreases as $n$ increases. Using the $l=2$-mode excitation energy $\Delta E^{int}_n=\frac{1}{2}M\omega_{n2}^2R^2$, the sum in Eq.~(\ref{k21}) can be approximated as
 \begin{gather}
k_2~\simeq~\dfrac{3}{4R^5}\dfrac{Q_1^2}{\Delta E^{int}_1}~.
\label{kr}
 \end{gather}

We now observe that a similar expression to the one in Eq.~(\ref{kr}), appears in  Eq.~(\ref{phasemain}) which determines the phase shift $\Delta \Phi_{122}$. This allows to express $\Delta \Phi_{122}$ in terms of $k_2$,
		\begin{gather}
			\Delta \Phi_{Res}~=~\dfrac{25\pi }{1536}\dfrac{k_2}{ q(1+q)}~.
			\label{phase2}
		\end{gather}
We are interested in the case of small mass ratios, $q\lesssim 1/1000$ and a small but not extremely small $k_2$, $k_2 \lesssim 1/10$. Then we can parameterize the resonance dephasing by
\begin{gather}
		\Delta \Phi_{Res}~\simeq~ 5\times \left(\dfrac{k_2}{10^{-1}}\right)\left(\dfrac{q}{10^{-3}}\right)^{-1}~.
		\label{phase2n}
\end{gather}
	The resonance-induced dephasing is governed by the dimensionless tidal Love number and the companion's mass ratio. Generally, the detection threshold for the instantaneous phase jump requires $\Delta \Phi_{Res} \gtrsim 1$ \cite{Lindblom:2008cm}. Thus, for typical values of Love numbers $k_2 \lesssim 10^{-1}$, it is more likely to observe resonances for moderate to extreme mass-ratio binaries $10^{-3}\leq q \leq10^{-5}$.

We can also express $k_2$ in terms of the frequency $\omega_{12}\equiv\omega_2$ of the $n=1$, $l=2$ mode. At resonance, from Eq.~(\ref{qnl}),  $Q \sim \Delta E^{int}$, where  $\Delta E^{int}=\frac{1}{2}M\omega_2^2 R^2$ is the energy of the oscillating star at resonance.
Thus, on dimensional grounds, we get 	$Q\sim\Delta E^{int}R^2 $.
For example, for a constant energy density perturbation $Q=\frac{3}{5}\Delta E^{int}R^2 $, while typical non-constant energy density profiles result in a numerical prefactor $\lesssim 1$ \cite{Andersson:2019ahb} (see also \cite{qLove}).
Substituting the expressions for $Q$ and $\Delta E^{int}$,  we arrive at our final result for the Love number
\begin{gather}
	k_2~\simeq~\mathcal{N}\omega_{2}^2R^2~,
	\label{lovemain}
\end{gather}
where $\mathcal{N}$ is an order unity dimensionless number that depends on the object's energy density profile and contains the numerical factors in the definition of the Love number \cite{qLove}.
We will use Eq.~(\ref{lovemain}) to determine the detectability of $k_2$ in the next section.

Remarkably,  in \cite{qLove}, it is shown that the gravitational polarizability of objects which possess a discrete spectrum of quantum mechanical energy levels is similar to that of classical stars.  This follows from the fact that the wavelength of the oscillation is of order of the star radius. We shall refer these objects as “quantum black holes” (QBHs) to mean the quantum state that corresponds to a classical BH. The idea is justified on the grounds of the Bohr correspondence principle, where at macroscopic excitations, expectation values are replaced by classical observables. Therefore, an excited quantum macroscopic object can be treated as a semi-classical oscillating fluid-like object that satisfy Eq.~(\ref{eom1}). Using standard time-independent quantum perturbation theory, the Love number of QBHs is given by \cite{qLove,cLove}
   \begin{gather}
   	k_2 \simeq \dfrac{3}{4R^5}\dfrac{|\langle \Psi_0|\hat{Q}|n=1,l=2\rangle|^2}{\Delta E_{1}^{int}}~.  \label{ql2} \end{gather}
 where $\Psi_0$ is the QBH ground state, $\hat{Q}$ is the mass moment operator that obeys the no-hair theorem; $\langle \Psi_0|\hat{Q}|\Psi_0\rangle=0$. The definition of Eq.~(\ref{kr}) is restored by applying the Bohr correspondence principle and replacing expectations values with classical observables, $\langle \Psi_0|\hat{Q}|n,l=2\rangle \leftrightarrow Q_n$. In this form, Eq.~(\ref{ql2}) can be treated in a similar way to the classical treatment of Eqs.~(\ref{kr}),(\ref{lovemain}), which eventually recovers the result
$ 	k_2\simeq\mathcal{N}\omega_{2}^2R^2$. The result is valid for any object of radius $R$, quantum or classical, which has a quadrupole internal mode whose non-relativistic frequency is $\omega_2$.

\section{Detectability}

In this section, using the Fisher method, we give a quantitative estimation of the statistical error in measuring the Love number. We  discuss the prospects for detection of a non-vanishing Love number with the future space LISA detector and demonstrate that during the inspiral, it is more likely to detect the Love number with resonances rather than tidal deformability.

We evaluate the detectability of the Love numbers through resonant excitations with the planned space telescope LISA, which according to \cite{LISA:2017pwj}, could track and observe moderate to extreme mass-ratio binaries from the early stages of the inspiral and up to the merger with high SNR.
Before addressing the precise statistical analysis, we wish to emphasize that for most of the range of the binary masses and spins and for Love numbers $k_2 \lesssim 10^{-1}$, the leading order 2.5PN resonance phase term is comparable to the other effects entering at 2.5PN, such as  the PP 2.5PN term and the leading order tidal heating term. For smaller values of $q$, the resonance phase term becomes  significant.  Since it is established that LISA can detect the other 2.5PN effects, we expect that LISA could  be able to detect the Love numbers with high confidence.
	
To evaluate the statistical error, we employ the Fisher information method. Assuming a signal $s(t)=h(t,\theta^i)+n$, with the uncorrelated noise $n$, a model signal $h(t,\theta^i)$ with model parameters $\theta^i$.
For high SNR events, the posterior distribution takes the form
		\begin{gather}
			p(\theta^i|s)\propto e^{-\frac{1}{2}\Delta \theta^i \Delta \theta^j \Gamma_{ij}}~.
		\end{gather}
		where $\Gamma_{ij}$ is the Fisher matrix defined as
		\begin{gather}
			\Gamma_{ij}~=~\left(\dfrac{\partial h}{\partial \theta^i}\Big|\dfrac{\partial h}{\partial \theta^j}\right)~.\label{FisherM}
		\end{gather}
		with the inner product defined by $(h_1|h_2)=4\text{Re}\int_{f_{min}}^{f_{max}}\frac{\tilde{h}_1(f)\tilde{h}^*_2(f)}{S_n(f)}df$, and $S_n(f)$ is LISA's design noise spectral density. We choose $f_{max}=f_{ISCO}(\chi)$, where $f_{ISCO}$ is the orbital frequency at the innermost stable circular orbit(ISCO) and $f_{min}=10^{-5}$Hz as the lowest frequency in the LISA frequency band. The model parameters are $\theta^i=(\ln \mathcal{A}, \ln \mathcal{M}_c, \eta,\Phi_c, t_c, \chi_1, \chi_2, k_2)$, where $\mathcal{A}$ is the amplitude, $\mathcal{M}_c$ is the chirp mass, $\eta$ is the symmetric mass-ratio, $\Phi_c$ and $t_c$ are the phase and time at coalescence, $\chi_i$ are the companions spin parameter and $k_2$ is the Love number given in Eq.~(\ref{lovemain}). The statistical error in measuring $k_2$ is related to the Fisher matrix,
		\begin{gather}
			\sigma_{k_2}~=~\sqrt{\langle \Delta k_2\rangle^2}~=~\sqrt{(\Gamma^{-1})_{k_2 k_2}}
		\end{gather}

		We consider quasi-circular orbits and employ the analytical frequency domain post-Newtonian approximation \texttt{TaylorF2}, which accurately describes the binary evolution of the inspiral up to the ISCO \cite{Damour:2000zb,Arun:2004hn,Buonanno:2009zt}.
		The frequency domain GW waveform describing the binary inspiral is of the form  $\tilde{h}(f,\theta_i)~=~\mathcal{A}e^{i\Phi}$, where $\Phi$ is the phase evolution in Eq. (\ref{pf}). From Eq.~(\ref{lovemain}),  for $q\ll1$, the instantaneous phase shift at resonance becomes
		\begin{gather}
			\Delta \Phi_{Res}~\approx~\mathcal{N}~\dfrac{\omega_2^2R^2}{20q}.
		\end{gather}
In our analysis we included correction terms up to 3PN order and neglected the higher order tidal deformability terms that depend on the Love number and enter at 5PN and 6PN order (See Sec.~\ref{tdres}).
		
		\begin{figure}[t!]
			\hspace{-1.75cm}
			\includegraphics[width=1.1\linewidth]{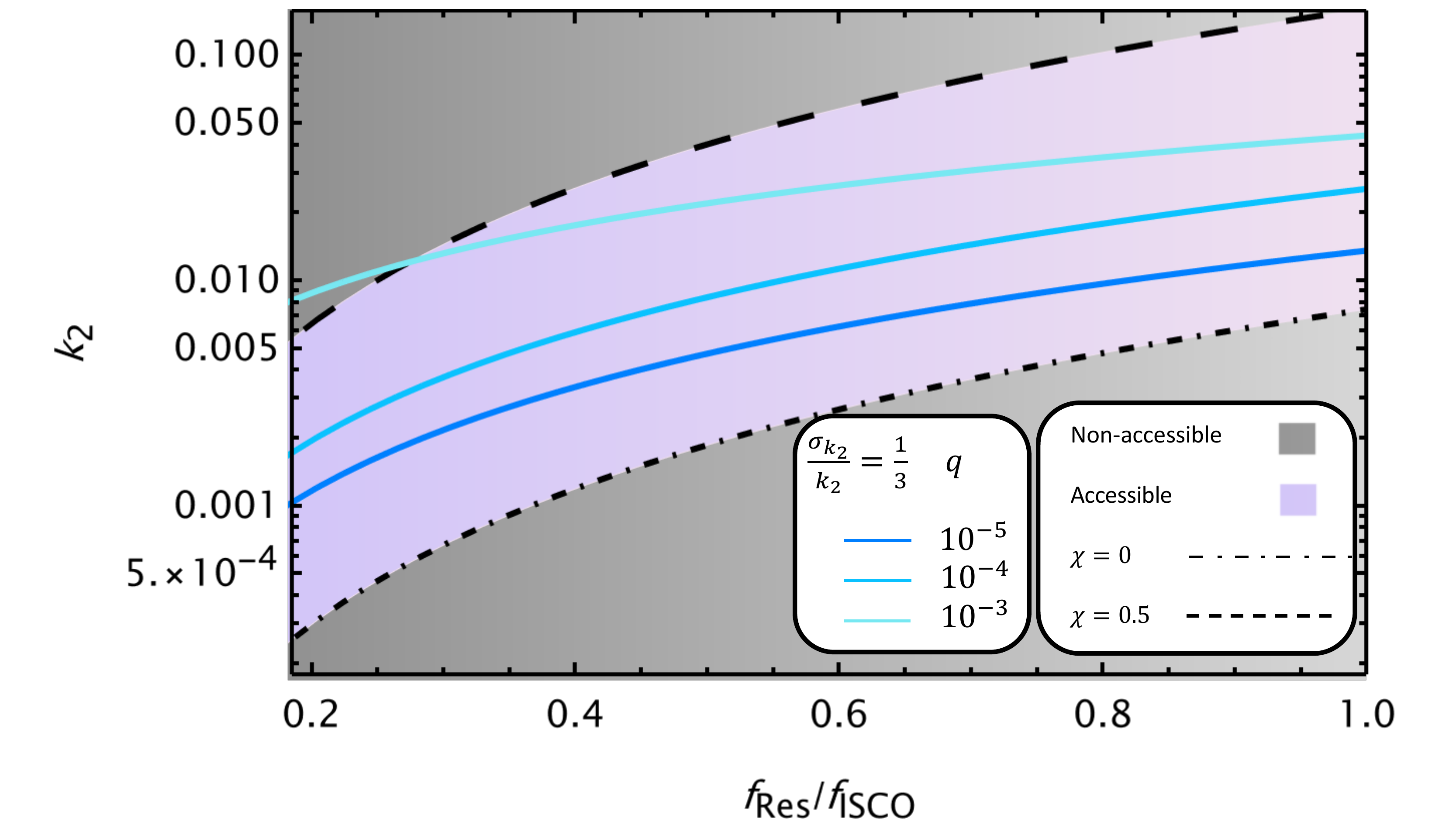}
\caption{{\small The solid blue lines correspond to a potential  measurement of $k_2$ for a given mass ratio $q$, with relative error $\sigma_{k_2}/k_2=1/3$. The region above each solid line corresponds potential measurement of $k_2$ with a relative error smaller than $1/3$. As anticipated by Eq.~(\ref{phase2}), for a smaller mass ratio, the error on measuring a specific $k_2$ is smaller and it is possible to measure smaller values of $k_2$.   The purple region describes the parameter space accessible to our model for values of the spin parameter between 0 and 0.5, taking into account Love-resonance-spin relation: $k_2\propto \omega_{ISCO}(\chi)R(\chi)$, such that a given Love number corresponds to a specific resonance frequency and spin parameter. The gray region describes the parameter space which is not accessible to our model for these values of the spin parameters. } }
\label{s2}
\end{figure}
		
Additionally, since our model is valid only until the ISCO, the frequency range $\omega_{2}>\omega_{ISCO}$ is not included in our analysis. Consequently, it is beneficial to parameterize the oscillation frequencies in terms of the ISCO frequency $\omega_{2}=\alpha \omega_{ISCO}$, where $0<\alpha\leq1$, and $\omega_{ISCO}(\chi)$ is spin-dependent. This also means that resonance at the ISCO sets the maximal value of the Love number that can be detected $k_2^{max}=\mathcal{N}\omega_{ISCO}^2R^2$.

	We consider moderate to extreme mass-ratio binaries with $q= [10^{-3},10^{-4},10^{-5}]$, where the central object mass is $M_1=10^6M_{\odot}$, and small to moderate Kerr spin parameters $\chi^i=[0, 0.1, 0.2, 0.3, 0.4, 0.5]$, at a luminosity distance $D_l=2$Gpc. We also average over the sky location parameters \cite{Buonanno:2009zt}. We assume equal spins $\chi_1=\chi_2$ that are aligned with the orbital angular velocity vector.   For the model-dependent order unity coefficient $\mathcal{N}$, we use the estimation derived in \cite{qLove}, and consider $\mathcal{N}\in[0.1,1]$.

In Fig.\ref{s2}, the purple region shows the analytical Love-resonance-spin relation described in Eq.~(\ref{lovemain}) that is determined by our model, where a given Love number corresponds to a specific resonance frequency and a spin parameter. This region describes the parameter space accessible to our model and is independent of the detector properties. In our analysis, the largest accessible $k_2$ is reached for $\mathcal{N}=1$, $\alpha=1$ and $\chi=0.5$, resulting in $k_2^{max}\approx0.159$, larger values are inaccessible to our model.
The gray region is the parameter space region that our model cannot describe.

\subsection{Comparison to Tidal-deformability}\label{tdres}

We now turn to estimate the relative magnitude of the resonance phase shift effects compared to the magnitude of tidal deformation effects on the phase evolution.  To leading PN order, the  tidal deformability contribution to the phase for $q\ll1$ takes the form $\Phi_{TD}(f)\sim k_2v^{5}/q$, where $v=(\pi M f)^{1/3}$ is the orbital velocity. The accumulated phase  throughout the inspiral is given by
\begin{gather}
	\Delta \Phi_{TD}~=~ \int_{f_{min}}^{f_{ISCO}} f \dfrac{d^2 \Phi_{TD}(f)}{df^2}df\sim \dfrac{k_2}{q}v_{ISCO}^{5}~.
\end{gather}
	\begin{figure}[h!]\vspace{-1cm}
	\hspace{-0.8cm}
	\includegraphics[width=1\linewidth]{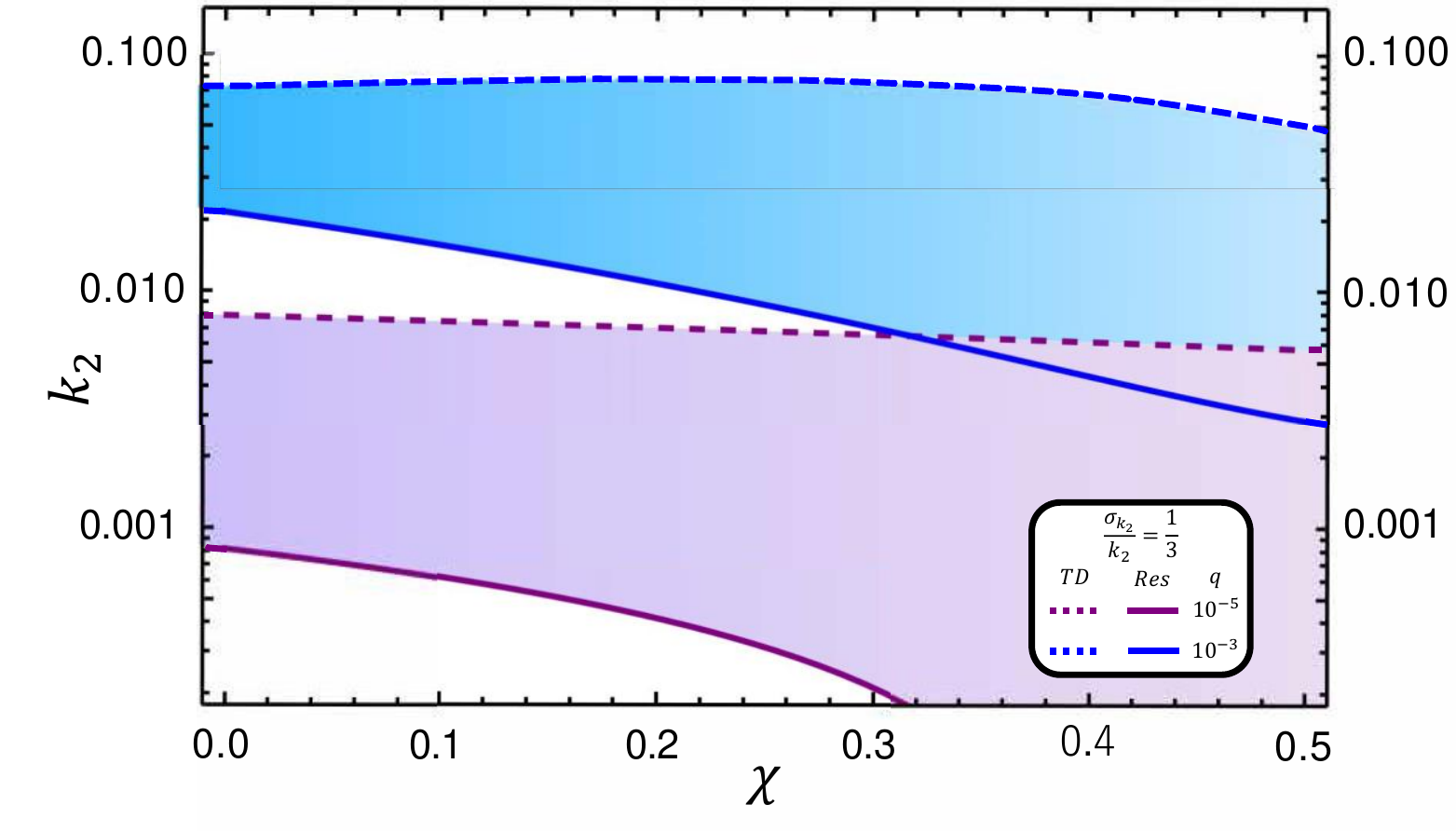}
	\caption{\small{The figure displays the relative statistical errors in measuring $k_2$ with resonance excitations and tidal deformation. For a given $k_2$ and $\chi$ we calculate $\sigma_{k_2}^{\text{res}}$ without tidal deformation effects and $\sigma_{k_2}^{TD}$ without resonances. The results show a preference for detecting $k_2$ with resonance effects. The preference is more apparent for a smaller  mass ratio $q$. The colored regions enclosed by the solid and the dashed lines mark the additional parameter space that resonances can probe compared to tidal deformation.}
	}
	\label{TDvsRES1}
\end{figure}

For a case for which the central object mass is $M_1=10^6 M_{\odot}$ and for small to moderate spin parameters, we find $v_{ISCO}^5\sim 0.01$. Comparing to the instantaneous resonance phase jump Eq.~(\ref{phasemain}), $\Delta \Phi_{TD}/\Delta \Phi_{Res}\sim v_{ISCO}^5$. Therefore, we would expect to have a larger error in the measurement of the Love number relying on tidal deformability.

We calculated the statistical error in measuring the Love number through tidal deformability and compared it to a measurement via resonance effects and found that the previous estimate is indeed correct. We repeated the statistical evaluation performed above, excluding resonance effects and including the leading tidal deformation terms entering the phase at 5PN  and 6PN order \cite{Bini:2012gu,Hotokezaka:2016bzh,Sennett:2017etc}.

The results  of the calculation of the ratio of the relative errors in measuring the Love numbers, denoted by $\sigma^{Res}_{{k_2}}/\sigma^{TD}_{{k_2}}$ for different spin parameters $0\leq \chi \leq 0.5$ are presented in Fig.~\ref{TDvsRES1}.

\section{Summary and conclusion}

The future measurement of GWs produced during BBH inspirals by the planned GW detector LISA will present an unprecedented opportunity to test GR. Hypothetical tidal interactions between the inspiraling objects would affect the waveform of the emitted GWs in a way that could only be possible if astrophysical BHs were actually ultra-compact objects possesing an internal structure rather than the structureless objects predicted by GR.

We discussed how the resonant excitation of the hypothetical non-relativistic interior modes of astrophysical BHs changes the phase of the emitted GW waveform when compared to the phase predicted by GR.  The non-relativistic nature of the modes was crucial to the possibility of resonantly exciting them, because in this case they could be excited when the two objects are still far apart. In this case, the resonance occurs a long time before the ISCO is reached and leads to a significant dephasing. We find that regardless the specific details of the primary's interior composition, the phase shift is governed by a single intrinsic quantity - the dimensionless tidal Love number $k_2$.

We evaluated the statistical error in measuring the Love number $k_2$ by LISA
using the resonance effect. We concluded that the smallness of the resulting  statistical error indicates that $k_2$ could actually be detected by LISA with impressive accuracy by observing intermediate and extreme mass-ratio inspirals. We compared the statistical error for detection of the Love number relying on tidal deformation effects with the error when using resonance effects and concluded that prospects of measuring $k_2$ using resonance effects are much better. The results reveal additional sensitivity-enhancement factors whose origin is the Love-resonance-spin relation. First, the statistical error in measuring the Love number reduces for BHs with higher spin, because for such BHs, the inspiral duration is longer. Second,  the statistical error in measuring the Love number reduces if the inspiral includes a range of higher orbital velocities, which could lead to excitation of higher internal frequencies, which, in turn, correspond to the BH having a larger Love number.

Our conclusion is that the effects of resonant excitation of astrophysical BHs during intermediate and extreme mass-ratio inspirals provide the best opportunity for putting bounds on, or detecting, the tidal Love number of astrophysical BHs and thus providing evidence of physics beyond GR. Nevertheless, we stress that the results of our statistical analysis should be viewed as preliminary estimates for the detection prospects. A comprehensive statistical treatment requires more accurate waveform modeling and should consider LISA's ability to track and discriminate several EMRIs simultaneously \cite{LISA:2017pwj}.

Our analysis is based on a general theoretical framework which only requires the existence of a set of non-relativistic internal modes, and does not require specifying the detailed properties of the central object. The entire dependence on the interior composition is parameterized in terms of the dimensionless tidal Love numbers. Therefore our results can be applied to a wide range of ultra-compact objects or BHs mimickers.

\section*{Acknowledgments}
We thank Vitor Cardoso, Tanja Hinderer and Ely Kovetz for  useful discussions.
The research is supported by the German Research Foundation through a German-Israeli Project Cooperation (DIP) grant ``Holography and the Swampland'' and by VATAT (Israel planning and budgeting committee) grant for supporting theoretical high energy physics.

\renewcommand{\baselinestretch}{1.15}

\end{document}